\documentclass[conference,compsoc]{IEEEtran}

\AtBeginDocument{%
  }

\usepackage{pifont}
\usepackage{todonotes}
\usepackage[htt]{hyphenat}
\usepackage[aboveskip=-2pt]{subcaption}
\usepackage{graphics}
\usepackage{graphicx}
\usepackage{url}

\usepackage{breakurl}
\usepackage[breaklinks]{hyperref}
\usepackage{multirow}
\usepackage{booktabs}
\usepackage[utf8]{inputenc}
\UseRawInputEncoding
\usepackage{float}

\usepackage{fancyhdr}
\usepackage{hyperref}
\usepackage{graphicx}

\begin{document}

\title{Targeted and Troublesome: Tracking and Advertising on Children's Websites\\\small{\textbf{\color{red}{{WARNING: Contains potentially NSFW images}}}}}

\author{\IEEEauthorblockN{Zahra Moti}
\IEEEauthorblockA{Radboud University}
\and
\IEEEauthorblockN{Asuman Senol}
\IEEEauthorblockA{imec-COSIC, KU Leuven}
\and
\IEEEauthorblockN{Hamid Bostani}
\IEEEauthorblockA{Radboud University}
\and
\IEEEauthorblockN{Frederik Zuiderveen Borgesius}
\IEEEauthorblockA{Radboud University}
\and
\IEEEauthorblockN{Veelasha Moonsamy}
\IEEEauthorblockA{Ruhr University Bochum}
\and
\IEEEauthorblockN{Arunesh Mathur}
\IEEEauthorblockA{Independent Researcher}
\and
\IEEEauthorblockN{Gunes Acar}
\IEEEauthorblockA{Radboud University}
}

\maketitle

\begin{abstract}

On the modern web, trackers and advertisers frequently construct and monetize users' detailed behavioral profiles without consent.
Despite various studies on web tracking mechanisms and advertisements, there has been no rigorous study focusing on websites targeted at children.
To address this gap, we present a measurement of tracking and (targeted) advertising on websites directed at children. Motivated by the lack of a comprehensive list of child-directed (i.e., targeted at children) websites, we first build a multilingual classifier based on web page titles and descriptions. Applying this classifier to over two million pages from the Common Crawl dataset, we compile a list of two thousand child-directed websites. Crawling these sites from five vantage points, we measure the prevalence of trackers, fingerprinting scripts, and advertisements. Our crawler detects ads displayed on child-directed websites and determines if ad targeting is enabled by scraping ad disclosure pages whenever available. Our results show that around 90\% of child-directed websites embed one or more trackers, and about 27\% contain targeted advertisements---a practice that should require verifiable parental consent. Next, we identify improper ads on child-directed websites by developing an ML pipeline that processes both images and text extracted from ads. The pipeline allows us to run semantic similarity queries for arbitrary search terms, revealing ads that promote services related to dating, weight loss, and mental health, as well as ads for sex toys and flirting chat services. Some of these ads feature repulsive, sexually-explicit and highly-inappropriate imagery. In summary, our findings indicate a trend of non-compliance with privacy regulations and troubling ad safety practices among many advertisers and child-directed websites. To ensure the protection of children and create a safer online environment, regulators and stakeholders must adopt and enforce more stringent measures.
\end{abstract}
\section{Introduction}
The proliferation of online tracking for analytics, behavioral advertising, and marketing has resulted in over a decade's worth of research into this ecosystem. Prior research has shown that not only is online tracking rampant on the web~\cite{englehardt2016online} but that trackers use increasingly-invasive tracking mechanisms---e.g., third-party cookies, tracking pixels, evercookies, and browser fingerprinting~\cite{ayenson2011flash,bielova2020missed,acar2014web,englehardt2016online}---to relentlessly build detailed profiles of users across the web without any consent for targeted advertising.

Such privacy concerns aside, online advertising has shown to be problematic in other ways. Ads and ad networks are a vector for distributing ransomware, malicious programs, and cryptojackers---posing a serious security threat to users~\cite{zarras2014dark,li2012knowing,subramani2020push, malware-google-ads, ads-ransomware,MalVirt,YouTube_serves_ads,bitdefender_blog,tekiner2021sok}.
Ad networks also suffer from click fraud, which is estimated to reach \$100 billion in 2023~\cite{wilbur2009click,ad_fraud_cost}. Finally, online ads often contain clickbait, untrustworthy, or distasteful content that peddle software downloads, listicles, and health supplements---all of which users find problematic to their online experience~\cite{zeng2021makes}. 

While online tracking and targeted advertising pose a threat to users of all ages, children especially bear an acute cost. Children may not fully understand the consequences of online tracking and revealing their personal data online~\cite{silly_name, telling_passcodes}, but they yield immense ``pester power'' to influence their parents' purchase decisions~\cite{lawlor2011pester}. Thus, children are an attractive target audience for advertisers and marketers alike~\cite{lawlor2011pester,kunkel2004report}, as they are more vulnerable to persuasive advertising~\cite{john1999consumer,buijzen2007reducing,cai2013online}, and susceptible to harmful content~\cite{rozendaal2010comparing,ali2009young}.

Despite the aforementioned evidence that suggests a differential impact on children, there is little empirical research on online tracking and advertising practices on children's websites.
The lack of a comprehensive and updated list of websites directed at children poses a major challenge for studying children's websites.
Previous large-scale internet measurement studies have relied on popular website lists such as Tranco~\cite{tranco_list} and Alexa~\cite{alexa-com-wayback} (before it shut down in 2021~\cite{alexa-retired}), but these lists may not specify website categories, and even when they do, the website categories may not be reliable and comprehensive~\cite{mathur2019dark,Vallina2020domainclassification}. As a result, prior work~\cite{vlajic2018online, cai2013online} has only examined online tracking on at most a hundred children's websites and has been restricted in scope and methods---lacking a comprehensive investigation of both online tracking and advertising.
To overcome this limitation, we built our own repository of child-directed websites. We trained a text-based classifier that detects children's websites using HTML metadata fields such as \texttt{<title>} and \texttt{<description>}.
The classifier is based on a pre-trained multilingual model that we fine-tuned for our binary classification task. Applying the classifier to the Common Crawl dataset~\cite{CC-data}, we compiled a list of 2K manually verified child-directed websites.

\begin{figure*}
    \centering        
    \includegraphics[width=0.99\textwidth]{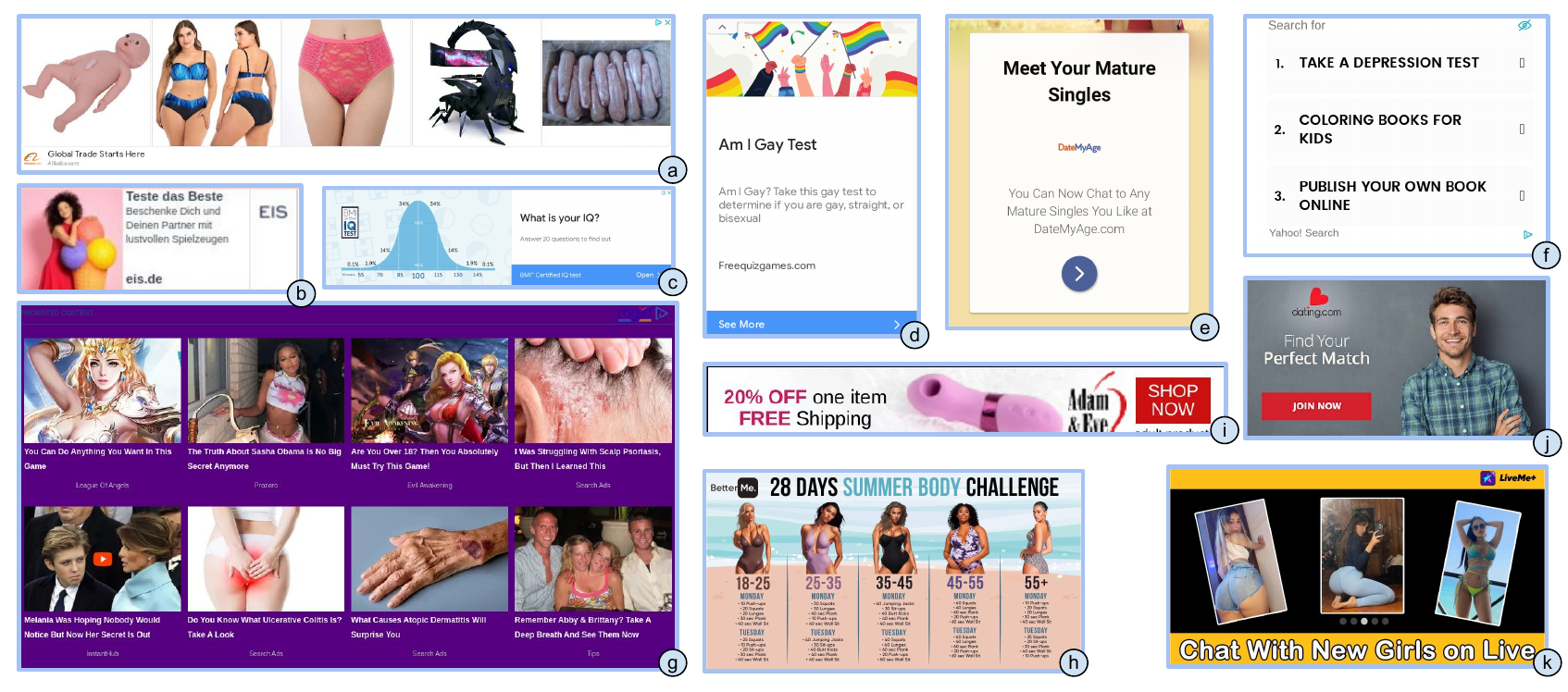}
    \caption{A sample of improper ads found on child-directed websites in our crawls.
    }
    \label{fig:bad-ads-collage}
\end{figure*}

To study several online tracking, ad targeting, and problematic ad practices, we crawl our list of 2K child-directed websites---varying the location (five vantage points) and form factors (desktop \& mobile). Starting with ad targeting, we study the extent to which ads that appear on children's websites are targeted---a practice that has come under increasing scrutiny both in the EU and the US~\cite{dsa_dma_package,sotu,coppa2}. We then present an exploratory analysis of ads from categories deemed problematic for children, such as dating, weight loss, mental health, and ads that contain racy content.
Next, we turn to online tracking, which is a necessary ingredient of behavioral advertising. We study the ecosystem of trackers and specifically quantify the prevalence of trackers, cookies, and use of browser fingerprinting techniques such as Canvas, Canvas Font, AudioContext Fingerprinting, and WebRTC local IP discovery~\cite{englehardt2016online}. 

Our work is especially pertinent in light of impending regulatory changes. In the US, there have been calls~\cite{coppa2} to update the Children's Online Privacy Protection Act (COPPA)~\cite{Coppa-FTC} in order to prohibit ``internet companies from collecting personal information from users who are 13 to 16 years old without their consent'' and to ``ban targeted advertising to children and teens.'' The current US President Joe Biden has called for a ban on collecting data on and serving targeted ads to children~\cite{sotu}; whereas in the EU, the upcoming Digital Services Act (DSA) will specifically prohibit ads targeted at children~\cite{dsa_dma_package}.

Our research seeks to offer empirical evidence on advertising and tracking practices employed on children's websites by making the following contributions:

\begin{itemize}
    \item Using a lightweight classifier based on web page metadata fields, we build a repository of child-directed websites and crawl them to measure tracking and advertising practices using multiple vantage points and form factors (desktop \& mobile).
    \item We measure targeted ads using two ad vendors' (Google and Criteo) ad disclosure statements, and find that targeting is enabled for 73\% of the ads we could measure.
    \item Using text and images extracted from the ads, we detect \textit{racy} ads, and ads about \textit{weight loss}, \textit{dating}, and \textit{mental health} using semantic similarity search based on a lightweight, multilingual language model. While this content analysis is exploratory, our method enables human-in-the-loop investigations with arbitrary queries, and it paves the way for the automatic content analysis of ads.
    \item We also find ads linking to malicious content, improper ads of sex toys, dating services, and ads containing sexually suggestive images (Figure~\ref{fig:bad-ads-collage}).
    
\end{itemize}

All the data and software from our study will be made available to researchers.\footnote{We share the list of identified child-directed websites and a sample of advertisement disclosures on~\textit{ ~\url{https://github.com/targeted-and-troublesome/}}.}

\section{Related Work}

\subsection{Web tracking measurements}

Over the past decade, several web privacy measurements have shown the scale and complexity of online tracking~\cite{ roesner2012detecting, englehardt2016online, iqbal2021fingerprinting, cassel2022omnicrawl,dambra2022sally}. Research on \textit{stateful} tracking has examined how unique tracking identifiers are stored on the client side~\cite{sanchez2017web} using cookies~\cite{kristol2000http,chen2021cookie}, localStorage~\cite{ayenson2011flash}, cache (ETags)~\cite{ayenson2011flash}, or other client-side storage mechanisms.
On the other hand, research on \textit{stateless} tracking has examined the use of fingerprinting, a mechanism that exploits differences in browsers and devices to obtain a likely unique identifier~\cite{mayer2012third}. Past research has shown that there are various fingerprinting vectors, including fonts, clock skew, GPUs, audio hardware, installed writing scripts and browser extensions, among others~\cite{eckersley2010unique,kohno2005remote,mowery2012pixel, englehardt2016online, cao2017cross, DRAWNAPART,laperdrix2021fingerprinting}.
Research on defense against fingerprinting has contributed methods to detect fingerprinting, tracking and advertising~\cite{sjosten2021essentialfp,acar2014web,englehardt2016online,iqbal2021fingerprinting, iqbal2020adgraph,siby2022webgraph}.
Our study borrows heuristics from prior work~\cite{englehardt2016online,iqbal2021fingerprinting} to detect fingerprinting scripts, and we use existing filter lists to identify trackers and advertisers.

\subsection{Tracking \& ads on children's media}
Motivated by the challenges posed by ads to children,
Cai and Zhao~\cite{cai2013online} manually labeled ads displayed on 117 children's websites. They found that 68\% of the websites featured ads, and less than half complied with COPPA. The authors also argued that children are unlikely to distinguish many ads from the website's original content.
Vlajic et al. \cite{vlajic2018online} investigated online tracking on twenty websites from Alexa's 
``Kids and Teens'' category~\cite{alexa-com-wayback} from two vantage points (EU \& US). The authors manually analyzed the HTTP headers and quantified hidden images (i.e., likely tracking pixels) loaded from ads and analytics domains. Compared to this past work, we study orders of magnitude more websites, follow more rigorous tracking measurement methods, and compare results across different vantage points. Additionally, we automatically detect targeted ads using ad disclosure pages and present an exploratory analysis of the content of ads that appear on children's websites.

Focusing on mobile platforms, Reyes et al.~\cite{reyes2018won} dynamically analyzed around 6,000 free children's Android apps and found that most apps violate COPPA due to their use of third-party SDKs.

\subsection{Improper and malicious ads}

A recent line of research has investigated the content of online ads. Zeng et al.~\cite{zeng2021makes} conducted a survey with 1,000 participants to determine the type of advertising content (e.g., chumboxes, clickbait, political, and low-quality content) that makes people dislike ads. In~\cite{zeng2020bad}, the same authors also studied problematic ads in the news and misinformation websites, where they found problematic ads served by native ad platforms.
In a study leading to the 2020 US elections, Zeng et al.~\cite{zeng2021polls} found that ads for misleading political polls that aim to hoover email addresses are widely used in online political advertising.
Subramani et al.~\cite{subramani2020push} studied the role of web push notifications in ad delivery, especially malicious ads. 
Through a large-scale study of malicious ads, Zarras et al.~\cite{zarras2014dark} showed that some ad exchanges are more prone to serving malicious ads due to inadequate detection.
Akgul et al.~\cite{akgul2022investigating} examined influencer VPN ads on YouTube and found advertisements disseminating misleading claims about online safety.
Ali et al.~\cite{ali2022all} measured how the distribution of potentially harmful ads on Facebook varies across users. Venkatadri et al.~\cite{venkatadri2019investigating} used Facebook's advertiser interface to study how Facebook obtains personal identifiers used in advertising.
In a concurrent work, Zhao et al.~\cite{zhao2023mobile} analyzed mobile ads aimed at children, uncovering inappropriate advertisements by certified mobile ad SDKs.
Medjkoune et al.~\cite{medjkoune2023marketing} showed that advertisers can target ads to children by placing their ads in children-focused videos---bypassing YouTube's age restrictions.

\subsection{Ad transparency}
In response to concerns about targeted advertising, ad networks and platforms have offered ad transparency interfaces that allow users to ascertain when and how they are being targeted. Andreou et al.~\cite{andreou2018investigating} investigated Facebook's ad explanations and found that they are often incomplete or misleading.
Researchers have also argued that ad networks should provide users with interpretable explanations and increase the visibility of disclosure mechanisms~\cite{eslami2018communicating}. 

Bin Musa and Nithyanand~\cite{musa2022atom} developed ATOM, a technique for determining data sharing between online trackers and advertisers. They used simulated personas to crawl websites, collect ad images, and conduct statistical analyses to identify correlations between tracker presence and advertiser behavior.
Liu et al.~\cite{liu2013adreveal} developed a framework called \textit{AdReveal} to investigate different mechanisms used in online targeted advertising. 
Vallina et al.~\cite{Vallina2023advanced} used statements found in Google's ad disclosure pages in their crowdsourced measurement of online behavioral advertisements.
To detect stealthy ads that aim to bypass adblockers,
Storey et al.~\cite{storey2017future} developed an extension that detects the AdChoices icon using perceptual hashing. While we considered applying Storey et al.'s method, we found URL-based detection of ad disclosure links (\S\ref{ad_detection}) to be more reliable and efficient. 

\subsection{Website categorization}
The majority of studies on web categorization have focused on text-based classifiers because most web content and metadata are text-based~\cite{song2021hierarchical,zhao2019design}.
Various studies used machine learning models such as BERT and recurrent neural networks to learn contextual representations and features of web pages using meta tags and body content~\cite{Gupta-Ensemble, zhao2019design, song2021hierarchical}.
Other researchers proposed image-based web page classification techniques using pre-trained convolutional neural networks and using Google image search results~\cite{aydos2020web, lopez2019visual}.
In our work, we built a lightweight classifier by fine-tuning an existing distilled language model and using text-based website metadata to detect child-directed websites. 

\vspace{-0.2cm} 
\section{Building a list of child-directed websites}

\begin{figure}[htp]
\centering
\includegraphics[width=8cm]{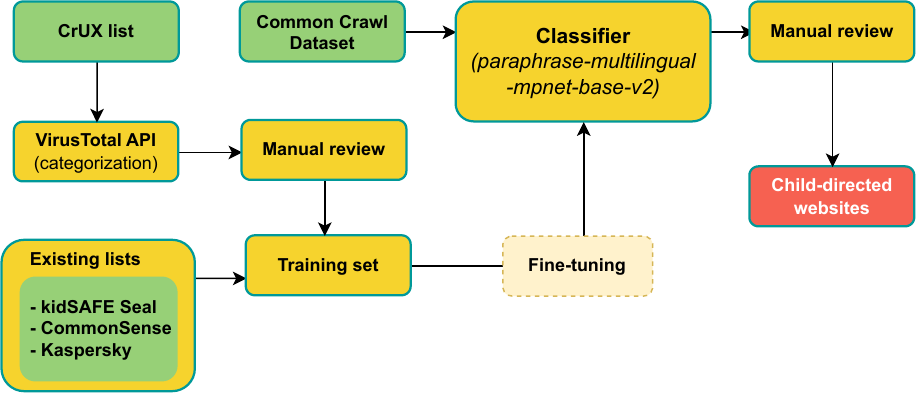}
\caption{Pipeline for building a list of child-directed websites.}
\label{fig:classifier}
\end{figure}
It is estimated that there are more than one billion websites on the Internet~\cite{num-of-websites-stats}, but only a small fraction are targeted at children. A central challenge, therefore, is identifying the websites that contain content directed to children. We initially searched and found three curated lists of children's websites: kidSAFE Seal Program~\cite{kidsafeseal}, CommonSense (filtered for children below the age of 13)~\cite{commonsense-list}, and a list compiled by Kaspersky~\cite{kaspersky-list}. Unfortunately, these lists contained only a total of 355 websites, some of which were no longer online.

To expand our limited list, we experimented with web categorization services such as McAfee, WebShrinker, and SimilarWeb, but decided to use VirusTotal's (academic) API because other services were either not freely available or did not let us query in bulk.
VirusTotal aggregates category labels from third-party scanners and categorization services, including BitDefender and TrendMicro~\cite{VT-domain-categorization-contributors}.
We used the VirusTotal API to retrieve web category data for the top one million websites from the Chrome User Experience Report (CrUX) list from May 2022~\cite{crux_urls}.
We observed VirusTotal's rate limits (20K/day/per academic license) during the process, which took roughly four weeks.
By searching for substrings ``kid'' and ``child'' in the returned category labels and removing false positives (such as ``Child abuse''), we obtained 1,264 websites categorized as related to children. However, our manual verification of these websites following the criteria presented in Appendix~\ref{criterion} revealed that 68.6\% of them were false positives, yielding only 396 child-directed websites.
Note that the low accuracy and inconsistency of domain classification/categorization services align with findings from prior work~\cite{Vallina2020domainclassification}.
Combining our initial 355 websites with our verified list of 396 websites and removing all inaccessible (5) and duplicate (164) websites, we obtained a total of 582 child-directed websites.

Motivated by the lack of accurate, up-to-date, and comprehensive sources of child-directed websites, we built a classifier to detect child-directed websites using the list of 582 websites as labeled data. Figure \ref{fig:classifier} illustrates the training and fine-tuning process. We define ``child-directed websites'' as websites that are primarily intended for use by children, and contain content, activities, or other features that are likely to be appealing to children (see Appendix ~\ref{criterion}, for our criteria).
Note that our labeling criteria do not fully overlap with COPPA's definition~\cite{Coppa-FTC}; for example, we do not require that sites have the actual knowledge of collecting children's data. Thus, we do not claim to measure compliance with COPPA or other relevant laws. In our study, children refer to individuals under 13, aligning with US and EU regulations (\S\ref{subsec:legal}).

\subsection{Labeled data for ML classifier}\label{data-prep}
Many web page classification methods use the entire text of the page \cite{Gupta-Ensemble} and its images~\cite{lopez2019visual}, which can be resource-intensive and time-consuming. Alternatively, researchers have explored web page classification on metadata fields such as \texttt{<title>}, \texttt{<description>}, and \texttt{<keywords>}, which tend to be shorter and shown to have strong correlations with the topic of web pages~\cite{song2021hierarchical}.
We followed the latter approach for its computational efficiency and reasonable accuracy.
Our preliminary analysis of over 500K web pages from the most popular one million websites in the Common Crawl dataset~\cite{CC-data} showed that more than 97\% of the websites have a title, 63\% of the websites include a description, and 24\% contain a keywords meta tag. Based on these availability statistics, we used the titles and descriptions for classification, leaving out the keywords.
To extract the titles and descriptions, we used the following HTML tags: \texttt{title, description, og:[title|description], and twitter:[title|description]}.
Applying this method to the WAT metadata files from the June-July 2022 Common Crawl snapshot~\cite{CC-data}, we extracted the titles and descriptions, limiting ourselves to the top million websites in the Tranco~\cite{tranco_list} or the CrUX~\cite{crux_urls} list.
We further refined our data by keeping a single page with the shortest URL from each hostname, which is more likely to be the home page. This resulted in metadata from 2.28 million pages, which included pages from the subdomains of the top million Tranco and CrUX websites.
We also extracted the same title and metadata information from the 582 known child-directed websites using a simple script based on Playwright~\cite{playwright}.
In both instances, when the page had more than one description or title available, we picked the longest one.

After completing the data collection process, we constructed a training set for our classifier. For negative samples, we randomly selected 2,500 of the 2.28 million pages and manually checked to remove children's websites. Our positive samples consisted of 576 title-description pairs after filtering out websites with titles shorter than ten characters.
\subsection{Building the ML classifier}\label{MpNet}
Our training data contained a limited number of labeled samples and our input consisted of text-based meta fields, potentially in multiple languages.
This made designing naive classifiers such as bag-of-words and TF-IDF less suitable for our task. Instead, we employed a pre-trained and multi-lingual language model.
Pre-trained models have proven to be adequate for general text classification tasks, but they need to be fine-tuned for more specific downstream tasks~\cite{Gupta-Ensemble}.
In particular, we decided to use the \textit{Paraphrase-Multilingual-MPNet-base-v2} (PM-MPNet-v2) model from the~\textit{SentenceTransformers}~\cite{reimers2019sentence,pm-mpnet-hugginface} library, which is a pre-trained multilingual and distilled model based on the MPNet method~\cite{song2020mpnet}. 
The distillation process~\cite{reimers2019sentence, distillation-process} involves training a smaller model (student) to mimic the behavior of a larger model (teacher). 
In particular, PM-MPNet-v2 is fine-tuned with a large number of paraphrased sentences in more than 50 languages~\cite{reimers2019sentence}.
However, PM-MPNet-v2 cannot be directly used for text classification since it only produces embeddings that are useful for semantic similarity-related tasks. Thus, we used HuggingFace's \textit{Trainer} API~\cite{trainer-api-hugginface} and \textit{AutoModelForSequenceClassification}~\cite{auto-models-hugginface} class to fine-tune the model and add a binary classification layer on top of the PM-MPNet-v2's embedding outputs. As input to the classifier, we used the concatenation of title and descriptions since this combination gave the best accuracy compared to using title or description alone.
In particular, we fine-tuned the model to detect child-directed websites using the training set explained in \S\ref{data-prep}.
We used the HuggingFace Transformers~\cite{hugging-hyperparameter} and Ray Tune libraries' Population Based Training (PBT) algorithm \cite{ray-tune,jaderberg2017population} to find the best-performing hyperparameters (batch size=12, epochs=2, and learning rate=4.2e-05).
The fine-tuning process took roughly five minutes on a consumer-grade GPU (GeForce RTX 3080 Ti).

To reduce false positives, we employed the modified ``Classify-Verify'' technique~\cite{stolerman2013classify}, which involves setting an acceptance threshold $t$, and accepting a prediction only if it is above $t$. 
Following Juarez et al. \cite{juarez2014critical}, we choose the threshold that maximizes 
$F_{\beta=0.5}$, which gives more weight to precision to reduce false positives. $F_{\beta}$ is a weighted harmonic mean of precision and recall, adjustable to emphasize either metric according to the classification task's needs~\cite{scikit-learn-fbeta}. A grid search of different threshold values shows that the maximum $F_{\beta=0.5}$ is achieved when $t = 0.93$, which reduces the false positive by $50\%$.
Ultimately, our classifier achieved a precision of 86\% and a recall of 70\% using 10-fold cross-validation, as detailed in Table~\ref{classify-verify-exp}.
\begin{table}[H]
\centering
\caption{Classification results before and after applying threshold (with 10-fold cross-validation).}
\begin{tabular}{lrrrrr} 
\hline
                  & \textbf{Precision} & \textbf{Recall} & \textbf{F-beta} & \textbf{TP} & \textbf{FP}  \\ 
\hline
Without threshold & 0.79               & 0.81            & 0.79            & 47          & 12           \\ 

With threshold    & 0.86               & 0.70            & 0.82            & 40          & 6            \\
\hline
\end{tabular}
\label{classify-verify-exp}
\end{table}

\subsection{The list of 2K children's websites}\label{}

Using the fine-tuned classifier, we calculated the label and probability score for 2.28M web pages from Common Crawl, excluding websites used in the training process. This process took roughly 5 hours. Our classifier identified 53,092 web pages as children's websites.
Due to time constraints, we focused on manually verifying the top 2,500 websites sorted by classifier probability, starting with websites that are most likely to be child-directed. 
An evaluation of our classifier and the details of our manual verification process can be found in Appendix~\ref{manual-verif}.

Our final list contained 2,004 websites in 48 distinct languages after eliminating false positives and deduplicating websites by their registrable domain (TLD+1). 
English was the most prevalent, accounting for 63\% of all websites. The prevalence of other prominent languages, including Russian, Spanish, French, German, and Portuguese, 
ranging between 3\% and 6\%.
The list included 582 websites from the training data and 1,422 websites identified by the classifier. 

\textbf{Website ranks:}
1,422 of the 2,004 websites were ranked in the top 1 million Tranco list (median rank 304K). 
While over a quarter of the websites are in the top 200K ranks, websites from all popularity levels are captured in our list. 
404 of the 582 websites that are not ranked by Tranco were ranked in the top one million by the CrUX list. Only
163 (8\%) websites were not ranked either by Crux or Tranco in the top one million.

\textbf{DNS0 Kids filter check:}
DNS0 Kids~\cite{DNS0} is a domain name resolver that detects and filters out content that is not suitable for children such as adult, dating, and copyright-infringing content. To find out the status of the websites in our list, we compared DNS0 Kids with CloudFlare's DNS resolver. If a website can be resolved by CloudFlare, but not by DNS0, we treated it as blocked. 
We found that only ten (0.5\%) of the 2,004 websites in our list were blocked by DNS0. Reviewing these ten websites, we found six of them to contain pirated videos, including cartoons. The remaining four websites contained activities for children but it was not obvious to us why they were blocked.
\section{Web Tracking and Advertising Measurements}

To assess the prevalence of trackers, fingerprinting scripts, and (targeted) advertisements on child-directed websites, we extended Tracker Radar Collector (TRC)~\cite{duckduckgo-trc}. TRC is a Puppeteer-based~\cite{puppeteer-gh} web crawler, which consists of modules called \textit{collectors} that record different types of data during a crawl, such as HTTP requests/responses, cookies, screenshots, and JavaScript API calls. New collector modules can be easily added to collect data necessary to perform different web measurements such as ours. Specifically, we added the following collectors to TRC:
\begin{itemize}
    \item ~\texttt{FingerprintCollector} (§\ref{fp_detection}): detects fingerprinting related function calls and property accesses
    \item ~\texttt{LinkCollector} (§\ref{inner_page_discovery}): extracts inner page links
    \item ~\texttt{VideoCollector} (§\ref{video_capturing}): captures the crawl video
    \item ~\texttt{AdCollector}  (§\ref{ad_detection}): detects ads and scrapes ad disclosures
\end{itemize}

We also used the existing TRC collectors, including \texttt{RequestCollector} to capture request/response details and detect tracking-related requests (§\ref{req_detection}), \texttt{TargetCollector} to detect newly-opened tabs in  §\ref{ad_detection},  \texttt{CookieCollector} to analyze cookies, and finally \texttt{CMPCollector} (§\ref{cmp-detection}) to interact with the consent dialogs and consent management platforms (CMP). 
We used TRC's anti-bot measures~\cite{duckduckgo-trc}, which thwarts bot detection to a certain extent by overwriting artifacts typically probed by anti-bot scripts (e.g., \texttt{navigator.plugins}, \texttt{Notification.permission})~\cite{TRC-antibot}. For the mobile crawls, we emulated a mobile browser using TRC's built-in features to spoof viewport dimensions, touch support, and the user-agent string.

\subsection{Identifying fingerprinting attempts} \label{fp_detection}
Identifying fingerprinting scripts can be challenging due to obfuscation and potential false positives. For example, scripts may use Canvas API for both drawing images or fingerprinting the user's browser~\cite{mowery2012pixel}. We draw on well-established methods to distinguish between fingerprinting and benign use of fingerprinting vectors~\cite{iqbal2021fingerprinting,englehardt2016online}. Specifically, we focused on Canvas, WebRTC, Canvas Font, or AudioContext fingerprinting and detected them using the heuristics presented by Iqbal et al.~\cite{iqbal2021fingerprinting}.
To detect fingerprinting attempts, we modified the \texttt{getter} and \texttt{setter} methods of the several Web APIs such as \texttt{CanvasRenderingContext2D.fillText} and 
\texttt{HTMLCanvasElement.toDataURL} to intercept potentially fingerprinting-related function calls and property accesses. Although TRC can intercept JavaScript API calls, we implemented a separate collector (\texttt{FingerprintCollector})
to avoid a known issue that prevented TRC from intercepting early function calls~\cite{ddg_issue}.
~\texttt{FingerprintCollector} simply injects the instrumentation script into each page and its descendant frames as soon as they are created.
We verified that our collector captures calls missed by TRC on both our custom-developed fingerprinting test pages and external demo pages such as BrowserLeaks~\cite{browser-leaks}.

\subsection{Identifying tracking-related requests}
\label{req_detection}
To identify tracking-related requests, we utilized the uBlock Origin Core~\cite{ubo-core} npm package, which mimics the tracking protection of the widely-used uBlock Origin extension~\cite{ublock}. We used uBlock Origin's default filter lists, which include EasyList and EasyPrivacy, among others~\cite{ubo-default-filterlists}. 
To accurately detect tracking-related requests, we provided uBlock Origin Core with the request's resource type (e.g., image or script) and the page and request URL derived from the HTTP request/response details recorded by the crawler.
Then, we mapped the tracker domains to their owner entities (i.e., organizations/companies) using DuckDuckGo's entity map~\cite{ddg_entities}. Using entities to quantify tracker prevalence reduces overcounting as multiple domains can be owned by the same business (e.g., googleanalytics.com and doubleclick.net are both owned by Google).

\subsection{Discovering inner pages}\label{inner_page_discovery}
We refrained from only focusing on homepages as prior work found that websites' inner pages tend to contain more trackers and cookies~\cite{urban2020beyond, aqeel2020landing}. Thus, we also gathered five inner links from each of the 2,004 websites by conducting four separate link-collection crawls (desktop and mobile crawls from Frankfurt and NYC). We preferred to crawl sites from two vantage points to reduce the time and effort required for the link collection process.
We excluded external domain links and documents such as PDFs and images, and we preferred links near the viewport's center to avoid unrelated links from footers or less visible page areas. After gathering these inner links, we merged them with the homepage URLs to form the final set.

\subsection{Interacting with consent dialogs}\label{cmp-detection}
Since the GDPR came into effect, websites typically show consent dialogs when viewed from the EU and to some extent even from the US ~\cite{rasaii2023exploring}. Ignoring these dialogs may lead to \textit{undermeasurement} of the tracking and advertising practices.
We decided to provide affirmative consent to all data processing request options (accept all) in our crawls to measure the full extent of advertisements and tracking a child could experience. To handle consent dialogs in an accurate and automated manner, we used DuckDuckGo's \texttt{autoconsent} library~\cite{duckduckgo-autoconsent}, which comes bundled with TRC~\cite{ddg-apps-autoconsent}. Autoconsent incorporates rules from Consent-O-Matic~\cite{consent-o-matic,nouwens2020dark}, allowing programmatic interactions with the CMPs.

\subsection{Video screen captures}\label{video_capturing}
To detect ads and scrape their disclosures, our crawler performed a series of interactions with the page, including dismissing popup dialogs, interacting with CMPs, and clicking on visible ad disclosure links(\S~\ref{ad_detection}). 
To monitor these interactions, we added a video capture functionality to the crawler (\texttt{VideoCollector}).
We used videos of the crawler's interactions to troubleshoot potential issues with the crawler process as well as to label animated ads and other crawl artifacts manually.

\subsection{Identifying ads and ad targeting criteria}
\label{ad_detection}

The \texttt{AdCollector} performed three main functions: 1) detecting ads, 2) scraping ads ---including its screenshots, links, iframes, scripts, videos, and background images, and 3) detecting and scraping ad disclosure pages to determine whether an ad is targeted or not.

\textbf{Detecting ads:}
To detect ads, we built on Zeng et al.'s~\cite{zeng2021makes} approach to use EasyList's rules~\cite{easy-list}.
EasyList rules are commonly employed by popular adblockers to block or hide ads. 
For each detected ad element, the crawler recorded a set of attributes, including its position on the page, its dimensions, class, ID, and name, in addition to the complete HTML source and a screenshot. If the ad element contained any child elements, which was mostly true, the crawler recursively recorded their details, including all links, images, video, script, and iframe elements. Small elements ($<30px$ in either dimension) and elements lacking any link, image, background image, or video were excluded.

The crawler not only took screenshots of each ad but also downloaded image and video elements that were descendants of the ad element. These media were utilized in the ML-based ad content analysis pipeline~\ref{analyzing_ad_content} alongside the ad screenshots. The crawler sent a single HTTP request during the page visit with the appropriate HTTP headers---such as the HTTP Referer [sic.] set to the current address-bar URL---when downloading these files. Finally, the crawler saved data-URL images found within the ad’s subtree.
Bin Musa and Nithyanand \cite{musa2022atom}  also utilized EasyList's rules for ad identification in their study on tracker-advertiser relationships, but their implementation differs from ours. While they focus on detecting image-containing HTTP responses using the EasyList filter set, we query the DOM to detect ad elements, such as div elements, and their relevant descendant elements, such as images, iframes, links (\texttt{a}), and videos. Operating at the DOM level also allows us to detect and scrape ad disclosure pages to detect targeted ads. 
To verify how accurately our crawler detects ads, we performed a sanity check on a random sample of 105 ads (15 ads from each crawl).
The crawler correctly detected ads in 85\% of cases, misidentified non-ads in 7.5\%, and captured blank or empty ads in 7.5\%. Some ad screenshots also included multiple (2.8\%) or only part (4.5\%) of the ads. However, the overall accuracy and quality of our ads appear to be higher than prior work by Zeng et al.~\cite{zeng2020bad}, which reported 34\% unrendered (blank/unreadable) ads. We attribute this difference in data quality to two potential reasons. First, we use a more realistic crawler equipped with anti-bot measures; and second, unlike Zeng et al., we opted not to click the ads---which may trigger more stringent anti-bot, anti-fraud protections that prevent the delivery or rendering of the ads.
Further, to evaluate whether our EasyList-based detector missed any ads, we manually reviewed 50 random pages where no ads were detected by the crawler. Our review did not reveal any false negatives, suggesting that our ad detection was robust.
We also verified the accuracy of the ad images separately downloaded by the crawler, finding all of them to be present in the ads shown on the page. 

\begin{figure}[h]
  \centering
  \begin{subfigure}[b]{0.5\textwidth}
    \includegraphics[width=\textwidth]{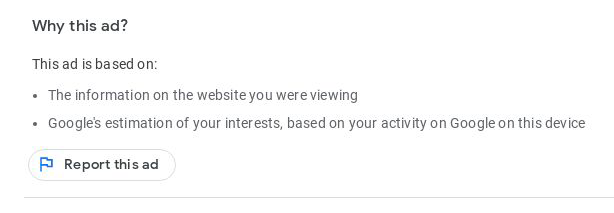}
    \caption{Targeted ad}
  \end{subfigure}
  \hspace{0.5cm}
  \begin{subfigure}[b]{0.5\textwidth}
    \includegraphics[width=\textwidth]{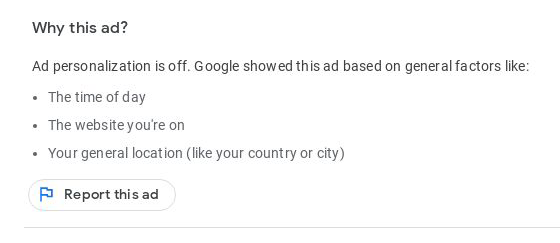}
    \caption{Non-targeted ad}
  \end{subfigure}
  \caption{Google's ad disclosure pages indicating whether an ad is targeted or not.
  The top figure belongs to a targeted ad 
  (indicated by \textit{Google's estimation of your interests}), 
  while the bottom one is for a non-targeted ad 
  (indicated by \textit{Ad personalization is turned off})
}
  \label{fig:google-ad-disclosure}
\end{figure}

\textbf{Determining targeting criteria:}
To measure the prevalence of targeted advertisements at scale, 
we automated the process of scraping ad disclosure (e.g., ``Why this ad'') pages.
While the content of ad disclosure pages may vary by ad platform, they generally explain in broad terms why a specific advertisement was shown to a user. The reasons may include, for instance, \textit{Google's estimation of your interests} or \textit{Websites you've visited}. The disclosure pages may also contain information about the website and the advertiser, and whether ad targeting is turned off for the website or a specific ad. Two example disclosure pages for a targeted and non-targeted ad are shown in Figure~\ref{fig:google-ad-disclosure}.

Ad disclosure pages are reachable by clicking the AdChoices icon \includegraphics[height=1em]{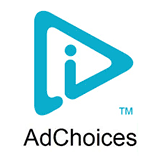} and the ``Why this ad'' button for Google ads~\cite{Google-ad-choices} and other ad providers. Initially, we attempted to detect the ad disclosure links using fuzzy image matching based on the AdChoices icon. However, we found that the icon's shape and visibility substantially vary across different ad vendors, and sometimes the icon can be hidden, making it unclickable. 
We chose to identify ad disclosure links through their URLs, focusing on a fixed set of providers we could detect reliably and deterministically. By analyzing ad disclosure pages in our pilot crawls, we compiled a list of hostnames (i.e., \texttt{adssettings.google.com}, \texttt{privacy.us.criteo.com} and \texttt{privacy.eu.criteo.com}) that appear in the ad disclosure links, and explain whether an ad is targeted or not. We limited our investigation to ad disclosure pages from these two providers because other providers we encountered in our pilot crawls did not offer useful information about the targeting criteria of the ads.

Once the crawler detects and clicks on the AdChoices link, the ad disclosure page opens in a new tab. We intercepted this new tab, stored its URL, screenshot and text contents (via ~\texttt{document.innerText)} for analysis. The scraped text contents are then used to detect whether ad targeting is enabled or not.
Specifically, we searched in the ad disclosure texts for specific disclosure statements indicating whether and how an ad was targeted. The disclosure statements include, for instance, \textit{Google's estimation of your interests} (targeted), \textit{Websites you've visited} (targeted) and \textit{Ad personalization is turned off} (non-targeted). If one or more statements indicating targeted ads occur in an ad disclosure text, we label the ad as targeted. Otherwise, we label the ad as non-targeted. Note that we count behavioral or retargeted ads also in the targeted category. We compiled a list of 18 statements (Appendix~\ref{ad_labeling}) incrementally, using over 40K ad disclosure texts extracted during the crawls.
We ensured that all ad disclosures contain at least one of these statements, to make sure our analysis is exhaustive.

\textbf{Interacting with the page and ads:} Upon loading a page, our crawler waited for 2.5 seconds and dismissed any popup dialogs using heuristics from prior work~\cite{mathur2019dark}. We dismissed these dialogs to prevent them from blocking our crawler's interactions with the webpage. The crawler then waited for another second before scrolling through the page in 10 steps, taking strides of about 500--600 pixels each interlaced with a random delay of 500--600 milliseconds. Finally, after waiting for another second, it scrolled up to the beginning of the page using the same scrolling behavior. We engineered this up-and-down scrolling behavior to allow the webpage to load any ad slots that are lazily loaded as the user scrolls the page below the landing fold.

The crawler then identified all ads on the page. It set the border color of each ad to red to visually mark the ads for manual review. 
The crawler then took a screenshot of the entire page and then scraped each ad in a top-down fashion. To ensure that an advertisement is fully seen, it scrolled down to each ad before taking its screenshot.
Finally, the crawler detected ad disclosure links and clicked each one individually to capture all ad disclosure texts and screenshots. 
We limited the number of scraped ads per page visit to ten, which limits over-representation by a few websites with many ads.

\textbf{Analyzing advertisement content:}\label{analyzing_ad_content} We identified and measured four kinds of ads in our corpus: weight loss ads, mental health ads, dating services ads, and ads that contain clickbait racy content. While our dataset of ads can be used to perform fuller content analysis, we focused on these four categories since prior work~\cite{gakdistressing,campbell2016rethinking} and regulatory reports~\cite{cap-code} have argued that these can be especially harmful to children. In fact, many ad networks' moderation policies~\cite{Google-kids-ad-policies, Taboola-restricted} explicitly restrict these categories of ads from appearing on children's websites. We note that the categories we focused on are not exhaustive, and our ads analysis is exploratory, serving as a preliminary investigation into this critical problem.

\begin{figure}[htp]
    \centering
    \includegraphics[width=8cm]{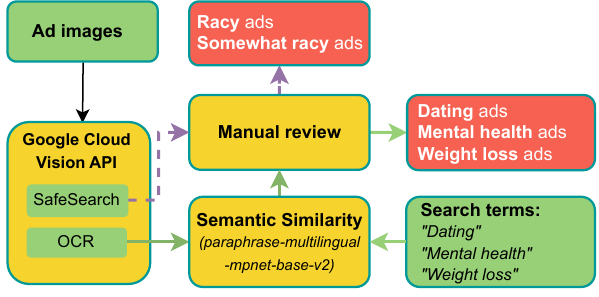}
    \caption{Overview of the ad content analysis pipeline.}
    \label{fig:ad_pipline}
\end{figure}

An overview of the ad content analysis pipeline is shown in Figure~\ref{fig:ad_pipline}.
To identify ads containing click-bait racy content, we employed
Google Cloud Vision API's SafeSearch Detection~\cite{google_cloud_vision_api}, which is a service that uses deep learning to analyze images and identify potentially unsafe content. It evaluates images against categories such as adult, violent, racy, and medical content and returns likelihood scores for each category, ranging from `VERY\_UNLIKELY' to `VERY\_LIKELY.'
Upon manually evaluating the output generated by the algorithm, we focused on the `racy' category with a likelihood of `VERY\_LIKELY'.
We also tested Microsoft's Adult Content Detection~\cite{microsoft2021adultcontent}, part of Azure Cognitive Services, to identify racy images. However, due to more false positives compared to Google Cloud Vision API, we chose the latter for our study.

We used the Google Cloud Vision API to extract text from ad images
following a similar approach to Bin Musa and Nithyanand~\cite{musa2022atom}.
The text in each image was extracted using the Optical Character Recognition (OCR) feature of the API, specifically by employing the \texttt{fullTextAnnotation} attribute of the API response. This allowed us to extract text data at different levels, such as page, paragraph, and word. We opted to use the paragraph level since it gives the best separation in ads promoting multiple unrelated products. Despite their name, paragraphs returned by the API were relatively short and akin to sentences (21 characters, on average).

We then employed semantic similarity to identify the most similar ad texts (paragraphs) corresponding to a given search query, which in our case were ``weight loss'', ``mental health'', and ``dating''. This approach is versatile and can be used to retrieve ads related to any arbitrary words or phrases. To compute the embeddings of the queries and ad paragraphs, we used the ``paraphrase-multilingual-mpnet-base-v2" model, the distilled multilingual model we used to classify web pages (§\ref{MpNet}). To find the most similar results, we calculated the cosine similarity between the embeddings of the search query and each ad paragraph and sorted them accordingly.
Next, we manually reviewed the 100 most similar distinct paragraphs and their associated images, including ad screenshots or background ad images, to identify those that pertained to the three categories of interest.
We also experimented with BERTopic \cite{bertopic} to create topic models and searched for clusters similar to our chosen categories. 
While this resulted in well-grouped texts, it required manual verification of numerous (several thousand) clusters. Sorting based on semantic similarity proved to be faster, more flexible, and easier to implement and evaluate, making it the preferred approach for manual reviewing.

\subsection{List of crawls}
\label{crawls-data}

The main dataset used in our study consists of seven crawls (Table \ref{table:crawl_stats}), all of which were run in April 2023 using cloud-based servers on Digital Ocean.
The crawls include five desktop and two mobile crawls from five and two vantage points, respectively. We limited the mobile browser crawls to two vantage points because we do not focus on mobile-desktop comparison, which we leave to future work. 
The vantage points used for crawls consisted of Frankfurt, Amsterdam, London, San Francisco, and New York City (NYC), to capture ads and tracking from different jurisdictions.
Crawls were run in parallel using separate servers with moderate resources (8 vCPU cores, 16GB RAM). Each crawl took between 13 and 32 hours to complete. 
A clean browser profile is used for each page visit to prevent ads targeted to our browsing history. During each crawl, we visited both landing and inner pages, following the process described in Section \ref{inner_page_discovery}.
When applicable, we accepted all personal data processing on consent (cookie) dialogs.
The order of visited pages was randomized within each vantage point. Note that the San Francisco crawl used inner links extracted from the NYC crawl, while the London and Amsterdam crawl used inner links from the Frankfurt crawl.
While this constraint did not appear to impact the success rate of visits across these vantage points, future research could explore identifying inner pages during the crawling process.

\section{Measurement Results}
Table \ref{table:crawl_stats} summarizes the overall statistics for measurement crawls.
A total of 71,567 pages were loaded successfully across all crawls. 
The success rate of our crawler was over 93\%, according to the successful visit criteria we developed and applied (Appendix~\ref{failed-visits}).
For simplicity, certain comparative results presented below are based on desktop crawls from NYC and Frankfurt, representing one location each in the US and the EU.

\begin{table}[]
\caption{Crawl statistics based on different vantage points. *: Avg./Sum of all sites visited across all crawls.}
\centering
\label{table:crawl_stats}
\begin{tabular}{llrr}
\hline
\textbf{\begin{tabular}[c]{@{}r@{}}Form\\ factor\end{tabular}} &
  \textbf{\begin{tabular}[c]{@{}r@{}}Vantage\\  point\end{tabular}} &
  \multicolumn{1}{l}{\textbf{\begin{tabular}[c]{@{}r@{}}Successfully \\ loaded pages\end{tabular}}} &
  \multicolumn{1}{l}{\textbf{\begin{tabular}[c]{@{}r@{}}Successful \\ crawling rate\end{tabular}}} \\ \hline
\multirow{5}{*}{\textbf{Desk.}} & NYC  & 10,310 & 95\%                       \\
                       & SF   & 10,301 & 95\%                       \\
                       & LON & 10,270 & 95\%                       \\
                       & FRA & 10,221 & 95\%                       \\
                       & AMS & 10,014  & 93\%                       \\ \hline
\multirow{2}{*}{\textbf{Mobile}} & NYC  & 10,168 & 94\%                       \\
                       & FRA & 10,283 & 96\%                       \\ \hline
\textbf{Avg./Sum}               &      & 71,567 & \multicolumn{1}{r}{95\%} \\ \hline
\end{tabular}
\end{table}

\subsection{Ad targeting and content analysis}
\label{ads-results}
Our crawler scraped 70,303 ads from 804 of the 2,004 distinct websites across seven crawls.
An average of 36\% of the pages contained one or more ads, and we detected targeted ads on 27\% of the pages we crawled.
The crawler scraped 10,839 and 9,447 ads on average in the crawls from the US and Europe, respectively.

\subsubsection{Over 70\% of ads with disclosures are targeted in nature}
\begin{figure}[]
    \centering
    \includegraphics[width=0.8\linewidth]{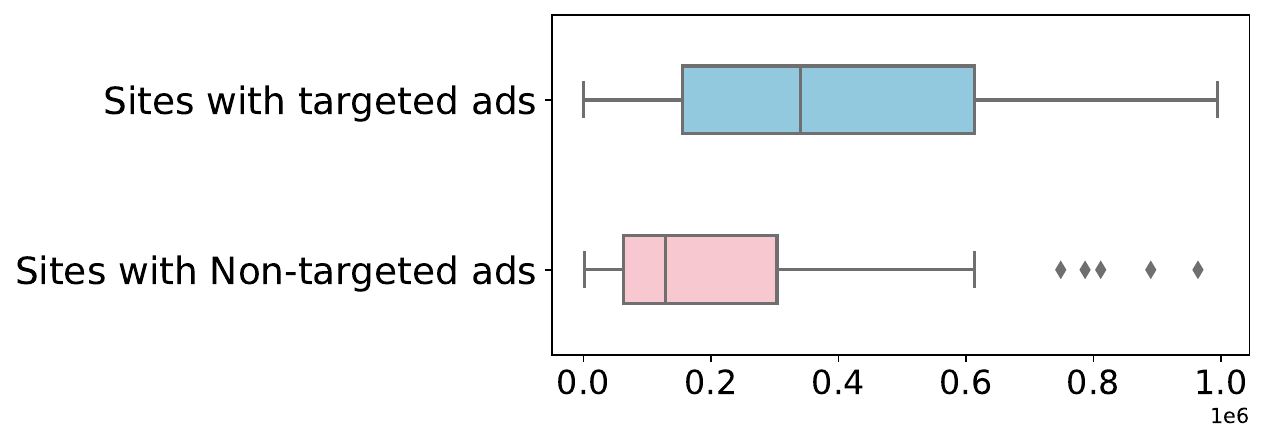}
    \caption{Tranco rank (x-axis) distribution of sites that use targeted vs. non-targeted ads. Popular websites (below) appear to be more prone to disabling ad targeting. }
    \label{fig:targeted_rank}
\end{figure}

Our crawler captured a total of 40,281 ad disclosure pages, which we used to determine the advertiser's identity and whether ad targeting is enabled or not.
There are fewer disclosure pages than ads due to ads without disclosure links and failures in detecting or opening those links.
In fact, we only consider ad disclosures from two ad providers: Google (97.8\%) and Criteo (2.2\%), since ad disclosure pages of other providers did not reveal the targeted status of the ad or the advertisers' identity.
Limiting our analysis to 40,281 ads with disclosure pages, we found that targeting was enabled for 73\% of the ads with disclosures.
Comparing across different privacy jurisdictions, we find 68\% of the ads on average were targeted in the EU crawls, compared to 76\% in the UK and the US crawls. Comparing the crawls from the two US cities (SF \& NYC), we find that 67\% of the ads were targeted in the SF desktop crawl, compared to 79\% and 82\% in the NYC-based desktop and mobile crawls, respectively. 
Although these variations might be attributed to stricter privacy regulations like CCPA and GDPR, our available data and methods do not permit us to make this attribution.
Comparing the Tranco ranks of the 689 websites that contain at least one targeted ad to 59 websites that only contain non-targeted ads, we find a tendency for popular websites to disable ad targeting (Figure~\ref{fig:targeted_rank}). Sites with targeted ads had a median rank of $\sim340K$, while those with only non-targeted ads had a median rank of $\sim128K$. Note that we only include 40,281 ads for which we can determine the targeted status in this analysis.

\begin{table}[]
\caption{Number of visits and scraped ads, along with percentages of ads/targeted ads per crawl. *: Percentage of targeted ads is only based on ads with disclosures. In the rightmost two columns, we include a site if we scraped at least one ad/targeted ad from one of its pages.}
\label{tab:ads-overview}
\resizebox{\columnwidth}{!}{%
\begin{tabular}{@{}llrrrr@{}}
\toprule
\textbf{\begin{tabular}[c]{@{}r@{}}Form\\ factor\end{tabular}} & \multicolumn{1}{c}{\textbf{\begin{tabular}[c]{@{}r@{}}Vantage\\ point\end{tabular}}} & \multicolumn{1}{c}{\textbf{\# ads}} & \multicolumn{1}{c}{\textbf{\begin{tabular}[c]{@{}r@{}}\% sites\\ with\\ ads\end{tabular}}} & \multicolumn{1}{c}{\textbf{\begin{tabular}[c]{@{}r@{}}\% sites with\\ targeted\\ ads\end{tabular}}} & \multicolumn{1}{c}{\textbf{\begin{tabular}[c]{@{}r@{}}\% targeted\\ ads\end{tabular}}} \\ \midrule
\textbf{Desk.}                                                 & NYC                                                                                  & 11,288                              & 38\%                                                                                       & 30\%                                                                                                & 79\%                                                                                   \\
                                                               & SF                                                                                   & 10,950                              & 38\%                                                                                       & 28\%                                                                                                & 67\%                                                                                   \\
                                                               & LON                                                                                  & 9,702                               & 36\%                                                                                       & 27\%                                                                                                & 76\%                                                                                   \\
                                                               & FRA                                                                                  & 9,700                               & 36\%                                                                                       & 26\%                                                                                                & 68\%                                                                                   \\
                                                               & AMS                                                                                  & 9,250                               & 35\%                                                                                       & 26\%                                                                                                & 67\%                                                                                   \\ \midrule
\multirow{2}{*}{\textbf{Mobile}}                               & NYC                                                                                  & 10,278                              & 36\%                                                                                       & 29\%                                                                                                & 82\%                                                                                   \\
                                                               & FRA                                                                                  & 9,135                               & 33\%                                                                                       & 26\%                                                                                                & 70\%                                                                                   \\ \midrule
\textbf{Avg.}/\textbf{Sum}                                                     & \multicolumn{1}{c}{}                                                                 & 70,303                              & 36\%                                                                                       & 27\%                                                                                                & 73\%                                                                                   \\ \bottomrule
\end{tabular}
}
\end{table}

\subsubsection{Ads can be targeted from anywhere}
The ``About the advertiser'' section in Google's ad disclosures shows the name and location (country) of the advertisers. This information is only available in 70\% of the ad disclosures in our dataset.
Extracting these fields from the ad disclosure texts, we identified 1,685 distinct advertisers from 81 different countries. Advertisers with the most ads in our data are displayed in Table~\ref{tab:top-advertisers}.
We note that due to the transient, targeted and localized nature of ad campaigns, the list in Table~\ref{tab:top-advertisers} may not represent the most common advertisers on child-directed websites in general. Further, in certain cases (e.g., Gloworld LLC and Marketism), an advertising or marketing agency is listed on the ad disclosure page instead of the company offering the advertised products or services.

The top ten advertisers are located in seven different countries and three continents. We observed that many of those advertisers are located far from our crawl vantage points, thus indicating that children visiting websites in our list can be targeted with ads from anywhere in the world. Reviewing a sample of 100 ads from each advertiser, we marked in the rightmost column their predominant ad theme. Five of the ten advertisers display ads for search results about various products---such as depression tests, belly fat reduction, senior meals, and electronic payments---on lesser-known search engines such as IngoSearch~\cite{IngoSearch}. Ads from Betterme~\cite{betterme}, a ``behavioral healthcare app'' with more than 100M installations, featured plans for weight loss, muscle gain, and intermittent fasting (e.g., Figure~\ref{fig:bad-ads-collage}~\textcircled{h}).\footnote{We note that Better.me's data sharing practices with third parties were investigated by Privacy International, but the company reportedly took corrective action\cite{PI-betterme}.}  Brain Metrics Initiative displays ads for IQ tests, an example for which is given in Figure~\ref{fig:bad-ads-collage}~\textcircled{c}. Alibaba Hong Kong, on the other hand, displays ads featuring racy and disturbing images of products sold on alibaba.com. For instance, the ad on the top left (\textcircled{a}) in Figure~\ref{fig:bad-ads-collage} features recurring images in Alibaba ads: a naked baby model (leftmost), rabbit meat (rightmost), and a semi-transparent underwear ad in the middle. We investigate similar racy clickbait ads and other improper ads in the following subsection.

\begin{table}[]
\caption{Top ten advertisers by the number of ads across all crawls.}
\label{tab:top-advertisers}
\resizebox{\columnwidth}{!}{%
\begin{tabular}{@{}llrrl@{}}
\toprule
\textbf{Advertiser} & \textbf{Location} & \textbf{\# ads} & \textbf{\begin{tabular}[c]{@{}r@{}}\%\\ targeted\end{tabular}} & \textbf{Type of ads} \\ \midrule
Vinden.nl B.V. & Netherlands & 4,707 & 86\% & Search results \\
EXPLORADS & Cyprus & 3,265 & 73\% & Search results \\
\begin{tabular}[c]{@{}l@{}}All Response \end{tabular} & UK & 2,453 & 68\% & Search results \\
Gloworld LLC & USA & 2,365 & 55\% & Online learning \\
\begin{tabular}[c]{@{}l@{}}Amomama M.\end{tabular} & Cyprus & 921 & 72\% & \begin{tabular}[c]{@{}l@{}}Workout muscle\\ gain, weight loss\end{tabular} \\
\begin{tabular}[c]{@{}l@{}}Media Quest \end{tabular} & UAE & 910 & 79\% & Search results \\
\begin{tabular}[c]{@{}l@{}}Brain Metrics I.\end{tabular} & Cyprus & 814 & 50\% & IQ tests \\
BetterMe & Cyprus & 731 & 85\% & Weight loss \\
Marketism & Israel & 645 & 49\% & Search results \\
\begin{tabular}[c]{@{}l@{}}Alibaba.com HK\end{tabular} & \begin{tabular}[c]{@{}l@{}}Hong Kong\end{tabular} & 541 & 86\% & \begin{tabular}[c]{@{}l@{}}Products sold\\ on Alibaba.com\end{tabular} \\ \bottomrule
\end{tabular}%
}
\end{table}

\subsubsection{Improper ads on child-directed sites}
\label{subsec:improper-ads-results}

\begin{table}[]
\caption{Number of improper ads identified for each crawl.}
\label{tab:improper-ads}
\resizebox{\columnwidth}{!}{%
\begin{tabular}{@{}llrrrrrr@{}}
\toprule
\textbf{\begin{tabular}[c]{@{}r@{}}Form\\ factor\end{tabular}} & \textbf{\begin{tabular}[c]{@{}r@{}}Vantage\\ point\end{tabular}} & \textbf{Dating} & \textbf{\begin{tabular}[c]{@{}r@{}}Mental\\ health\end{tabular}} & \textbf{\begin{tabular}[c]{@{}r@{}}Weight\\ loss\end{tabular}} & \textbf{Racy} & \textbf{\begin{tabular}[c]{@{}r@{}}Some-\\ what\\ racy\end{tabular}} & \textbf{Total} \\ \midrule
\multirow{5}{*}{\textbf{Desk.}} & NYC & 4 & 21 & 16 & 21 & 26 & 88 \\
 & SF & 7 & 9 & 15 & 6 & 25 & 62 \\
 & LON & 10 & 17 & 48 & 12 & 31 & 118 \\
 & FRA & 1 & 0 & 48 & 19 & 25 & 93 \\
 & AMS & 8 & 4 & 82 & 10 & 33 & 137 \\ \bottomrule
\multirow{2}{*}{\textbf{Mobile}} & NYC & 22 & 25 & 113 & 98 & 17 & 275 \\
 & FRA & 18 & 5 & 190 & 11 & 6 & 230 \\ \bottomrule
\textbf{Total} &  & 70 & 81 & 512 & 177 & 163 & 1003 \\ \bottomrule
\end{tabular}%
}
\label{improper-ads}
\end{table}

In total, our crawler collected 199,935 screenshots and images from the 70,303 scraped ads. After deduplicating the images, we queried the Cloud Vision API to obtain the category and OCR texts of the resulting 98,264 distinct images.
We manually reviewed 741 images classified as `VERY\_LIKELY' racy by the API. Separately, we reviewed 1,136 ad images with OCR text semantically most similar to our search terms (mental health, dating, and weight loss). Due to study limitations, we only examined the ads related to the top 100 distinct texts for each term. Since each distinct text may appear in multiple ads differently, we labeled the images separately and used videos captured by the crawler when the ad was animated or the ad screenshot was obscured.
Table \ref{improper-ads} shows the number of improper ads identified in each crawl, amounting to 1,003 across 311 distinct websites. A notable finding is the higher prevalence of such ads on mobile devices compared to desktops in general. 

\textbf{Racy images.} We found 177 racy ads and 163 somewhat racy ads, considered edge cases due to their potential inappropriateness for child-directed websites. These ads were identified across 80 distinct websites mostly ranked within the top one million according to the Tranco list, with a median rank of 426K. Figure~\ref{fig:bad-ads-collage}~\textcircled{a}, ~\textcircled{g}, ~\textcircled{k} are examples of some of these ads. Notably, the majority of the racy ads were encountered in mobile crawls; especially within the NYC crawl (98/177). Out of 177 racy ads, only 38 had disclosure pages. Among these,  targeting was enabled for 35 ads.

\textbf{Mental health.} 
By manually labeling 236 ad images, we identified 81 ads offering mental health services on 48 distinct websites. Examples of ads in this category contained ``take a depression test'' (Figure~\ref{fig:bad-ads-collage}~\textcircled{f}), ``online psychiatrists,'' ``how to get over depression,'' and a ``mental health chatbot which helps people with depression.'' We excluded false positives that were not mental health service offerings, such as ads for ``mental health counselor salaries,'' ``online psychology courses,'' and ``psychology books.'' 

\textbf{Dating.}  Manually labeling 231 ad images, we identified 70 dating platform ads on 48 distinct websites, most of which targeted mobile users. The ads promoted dating platforms such as ``dating.com,'' and ``Live Me,'' a live streaming app with ads featuring suggestive imagery (Figure~\ref{fig:bad-ads-collage},~\textcircled{j},~\textcircled{k}). 
Another ad for DateMyAge.com featured a call to ``[m]eet your mature singles'' (\textcircled{e}). False positives removed during the manual labeling of this category included ads for customer relationship tools, romantic holiday tours, and online appointment services.

\textbf{Weight loss.} 
We identified 512 weight loss-related ads (plans, apps, products) on 170 distinct websites by labeling 669 ad images. Notably, there was a higher number of weight loss ads on mobile devices, indicating campaigns targeting mobile users. Examples of text featured in these ads included ``intermittent fasting for weight loss,'' ``keto weight loss plan,'' and ``eating plan to lose weight'' (Figure~\ref{fig:bad-ads-collage}~\textcircled{h}).

In Figure \ref{fig:bad-ads-collage}, we provide additional examples of advertisements that are likely not suitable for children. Examples of these included an ad for a test called ``Am I Gay Test'' ~\textcircled{d}, for a sex toy ~\textcircled{i} and a sex toy shop~\textcircled{b}\footnote{Reportedly Germany's largest online adult retailer~\cite{eis-de-largest-retailer}} featuring an image of ice cream that could be appealing to children, and ads featuring clickbait and sexually suggestive images. The ads were found on websites related to K-12 e-learning, kids games, coloring books and worksheets, among others. The ads in Figure \ref{fig:bad-ads-collage} do not necessarily fit our four investigated categories but showcase the diversity of improper ads.

\textbf{Malicious ad links.}
Finally, we present an exploratory analysis of whether ads on child-directed websites link to malicious pages.
We submitted a sample of links extracted from the ad elements to the VirusTotal API in August 2023.
Specifically, we removed links with duplicate hostnames, and for Google ads, we extracted a direct link to the ad landing page using the `adurl' parameter~\cite{ad-url-param}.
While the overwhelming majority of the links were classified as benign, 149 of the nearly 3,940 scanned links were flagged as malicious or phishing by at least one scan engine.
Notably, the word ``taboola''  was mentioned in 78 of the 149 detected links as a URL parameter that seems to indicate the ad network (\texttt{network=taboola}).

\subsection{Tracking and fingerprinting analysis}

\begin{table}[]
\caption{Average number of third-party and tracker domains, and the prevalence of tracking and fingerprinting on child-directed websites from five vantage points.}
\label{table:main-results}
\resizebox{\columnwidth}{!}{%
\begin{tabular}{llrrrrrr}
\toprule
\textbf{\begin{tabular}[c]{@{}r@{}}Form\\ factor\end{tabular}} & \textbf{\begin{tabular}[c]{@{}r@{}}Vantage\\  point\end{tabular}} & \multicolumn{1}{l}{\textbf{\begin{tabular}[c]{@{}r@{}}3rd-Party \\ domains\end{tabular}}} & \textbf{\begin{tabular}[c]{@{}r@{}}Tracker\\ domains\end{tabular}} & \textbf{\begin{tabular}[c]{@{}r@{}}Tracker\\ entities\end{tabular}} & \textbf{\begin{tabular}[c]{@{}r@{}}\% sites\\ with\\  3rd \\Parties\end{tabular}} & \textbf{\begin{tabular}[c]{@{}r@{}}\% sites \\ with\\ trackers\end{tabular}} & \textbf{\begin{tabular}[c]{@{}r@{}}\% site\\ with \\ FP\end{tabular}} \\ \midrule
\multirow{5}{*}{\textbf{Desk.}} 
    & NYC & 31.6  & 23.4  & 20.0  & 95\%   & 90\%  & 9\%  \\
    & SF & 29.3    & 21.3  & 17.8  & 95\%  & 91\%  & 9\% \\
    & LON  & 21.3 & 14.3  & 10.6  & 96\%  & 91\%  & 7\%  \\
    & FRA  & 23.2 & 15.6  & 11.7  & 95\%  & 90\%  & 10\%  \\
    & AMS  & 21.4 & 14.3  & 10.6  & 93\%  & 89\%  & 7\%   \\ \midrule
\multirow{2}{*}{\textbf{Mobile}}  & NYC   & 29.8 & 21.8  & 18.4  & 95\%  & 91\%  & 9\%   \\
    & FRA  & 22.6 & 15.2  & 11.5 & 95\%  & 90\%  & 11\%  \\ \bottomrule
\end{tabular}%
}
\end{table}

Table \ref{table:main-results} shows the prevalence of third-party trackers detected across different crawls. We find that around 90\% of the websites have at least one tracker domain, and over 93\% embed at least one third-party domain.


 %

\begin{table}[!]
\centering
\caption{Prevalence of tracker entities in terms of number of distinct websites in Frankfurt and NYC desktop crawls.}
\begin{tabular}{lrlr} 
\hline\multicolumn{2}{c}{\textbf { FRA }}  & \multicolumn{2}{c}{\textbf{ NYC }} \\
\hline  \textbf{Entity} & \textbf{\# Sites} & \textbf{Entity} & \textbf{\# Sites} \\
\hline   Google & 1,702 & Google &  1,718 \\
 Facebook & 458 & Microsoft & 549 \\
 Index Exchange & 424 & Adobe & 543 \\
 Xandr & 416 & Xandr & 516 \\
 Adform & 412 & The Trade Desk & 501 \\
 The Trade Desk & 390 & Index Exchange & 495 \\
 OpenX & 378 & IPONWEB & 467 \\
 Adobe & 366 & Facebook & 456 \\
 Quantcast & 361 & Magnite & 446 \\
 PubMatic & 359 & OpenX & 426 \\
\hline
\end{tabular}
\label{table:trackers_prevalence}
\end{table}

\textbf{Third-party trackers.}
\label{third-party-trackers} 
The average number of tracker domains per site differs significantly, e.g., 15.6 and 23.4 in Frankfurt and NYC crawls, respectively, while the median is 15 and 16, respectively. The difference in averages can be attributed to websites with the high number of trackers in the NYC crawl. This explanation is in line with the results displayed in
Table \ref{table:top_sites_trackers}, which shows the top five websites with the most trackers in Frankfurt and NYC crawls. Most of these websites are among the top one million, which means they likely receive substantial traffic. Notably, all of these sites displayed ads that were targeted. The numbers shown in the table - number of trackers, requests, and cookies - reflect averages across the web pages.
In the NYC crawl, visiting \texttt{mathfunworksheets.com} triggered a total of 1,547 requests involving 161 unique third-party tracker entities (i.e., organizations/companies).
Another website, \texttt{woojr.com} found to contain 148 distinct third-party tracker entities when visited from NYC. This website includes resources for children's activities and educational materials, including printable worksheets and fun activity pages.
When visited from Frankfurt, \texttt{www.wowescape.com}, a website offering various games for children and teenagers, triggered requests to 95 distinct third-party tracker entities.

\textbf{Most prevalent trackers.}
Table \ref{table:trackers_prevalence} shows the tracker entities with the most prevalence in Frankfurt and NYC desktop crawls. We found a tracking-related request to Google domains, including its analytics, advertising and tag management scripts on $\sim$84\% of the 2,004 child-directed websites in both crawls.
Facebook is the second most prevalent entity in the Frankfurt crawl mostly due to Facebook Pixel (on 427 websites), which facilitates ad retargeting and conversion measurement, among  others~\cite{facebook_pixel}. Largely thanks to 
Linked Insight Tag (\texttt{px.ads.linkedin.com}, 466 websites), Microsoft is the second most prevalent entity in the NYC crawl.
Linked Insight Tag serves multiple purposes, including retargeting, conversion measurement, and providing demographic insights about website visitors~\cite{linkdin_tag}. 

\textbf{Regional differences.}
To explore the differences in tracker entities across vantage points, we compared the tracker entities from Frankfurt and NYC desktop crawls.
Despite a considerable overlap among the detected tracker entities (Jaccard index=$0.85$), we also identified variations. Specifically, our investigation unveiled 47 tracker entities exclusive to the Frankfurt crawl and 118 tracker entities that were only found in the NYC crawl.
For instance, tracking related requests to \textit{advanced STORE}~\cite{advancedstore} (\texttt{ad4m.at} \& \texttt{ad4mat.net}, 236 websites)
exclusively appear in the crawl from Frankfurt, whereas \textit{Throtle}, a company that provides an identity graph to marketers and advertisers, only appears on 171 websites in the NYC crawl~\cite{Throtle}.
Furthermore, we find that the majority of the websites in both Frankfurt and NYC crawls (∼70\% and ∼72\%, respectively) contain third-party trackers that set at least one cookie with the \texttt{SameSite=None} attribute and a lifespan of over three months. Primarily through \texttt{doubleclick.net} domain, Google set these cookies on over 51\% of the websites. 
While identifying the individual purposes of these cookies is out of scope, this combination of cookie attributes (esp. setting \texttt{SameSite=None}) makes it possible to track users across websites.

\begin{figure}[]
    \centering
    \includegraphics[width=1\linewidth]{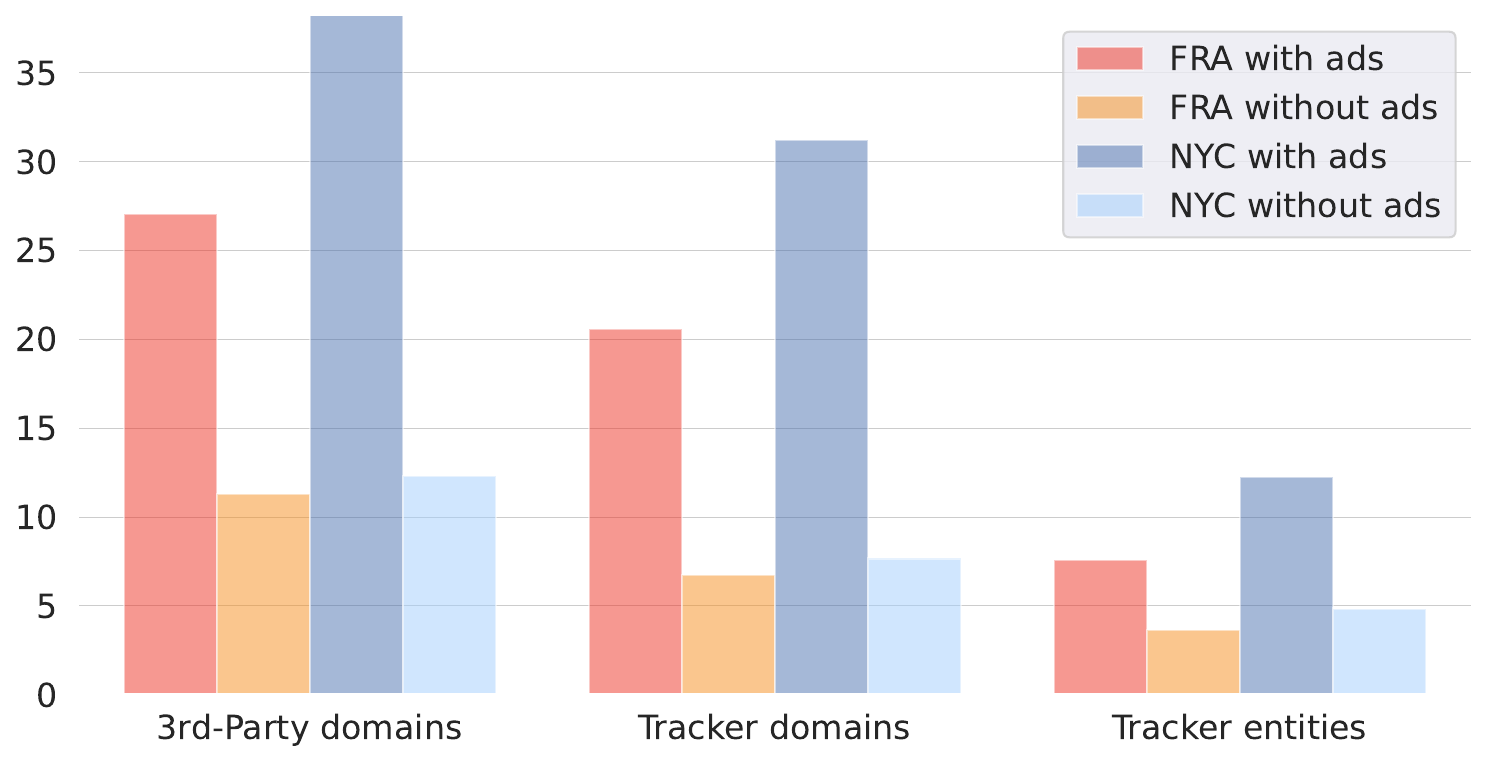}
    \caption{Comparison of the average number of third-party and tracker domains/entities on websites containing ads vs. not containing ads.}
    \label{fig:comp_analysis}
\end{figure}

\textbf{Sites with and without ads.}
As part of our investigation, we conducted an additional analysis to compare how the number of third parties and trackers change between websites with and without ads.
Figure~\ref{fig:comp_analysis} shows that websites with ads tend to
have substantially more third-party and tracker domains.
More specifically, the figure shows websites with ads tend to contain two to four times more third-party and tracker domains.

\begin{table}[]
\caption{websites with the most distinct tracker entities. The table shows the websites' distinct third-party tracker entities, the number of requests, cookies, and Tranco rank.}
\label{table:top_sites_trackers}
\resizebox{\columnwidth}{!}{%
\begin{tabular}{llrrrr}
\hline
\textbf{Loc.} &
  \textbf{Website} &
  \textbf{\begin{tabular}[c]{@{}r@{}}\# Trackers\end{tabular}} &
  \textbf{\begin{tabular}[c]{@{}r@{}}\# Requests\end{tabular}} &
  \textbf{\begin{tabular}[c]{@{}r@{}}\# Cookies\end{tabular}} &
  \textbf{Rank} \\ \hline
\multirow{5}{*}{\textbf{NYC}}  & mathfunworksheets.com       & 161 & 1,547 & 395 & 669K \\
                      & woojr.com                   & 148 & 2,181 & 391 & 83K  \\
                      & innerchildfun.com           & 139 & 1,235 & 336 & 308K \\
                      & kidzfeed.com                & 138 & 1,050 & 272 & 797K \\
                      & thecolor.com                & 138 & 1,068 & 260 & 192K \\ \hline
\multirow{5}{*}{\textbf{FRA}} & wowescape.com           & 95  & 392   & 55  & 258K \\
                      & webgames.io                 & 94  & 564   & 92  & 155K \\
                      & coloriages-pour-enfants.net & 90  & 401   & 66  & 919K \\
                      & theschoolrun.com            & 87  & 417   & 91  & 760K \\
                      & testsworld.net              & 86  & 478   & 138 & -    \\ \hline
\end{tabular}
}
\end{table}

\textbf{Browser Fingerprinting.}\label{subsec:fp-measurements} We now discuss our findings on fingerprinting scripts on child-directed websites.
Table \ref{table:main-results} shows that we detect fingerprinting scripts on 176 (9\%) and 218 (10\%) websites in Frankfurt and NYC crawls, respectively.
The overall prevalence of fingerprinting aligns with the recent research by Iqbal et al., which finds fingerprinting on 10.18\% of the top-100K websites~\cite{iqbal2021fingerprinting}.
One of the most prevalent fingerprinters in both crawls is an online payment company (Stripe; 66, 67 sites on Frankfurt and NYC crawls, respectively).
According to their help pages~\cite{stripe} Stripe primarily employs fingerprinting for fraud prevention purposes. 
Webgains (82 sites in the Frankfurt crawl), an affiliate marketing company, also mentions fingerprinting in their Data Processing Agreement with Merchants~\cite{webgains}, but without specifying its purpose.
The most commonly used fingerprinting method is Canvas fingerprinting, present on about 208 sites in the Frankfurt crawl and about 172 sites in the NYC crawl.
We found one or more trackers to be present on more than 90\% of mobile websites (Table \ref{table:main-results}), which is similar to our finding for the desktop websites.
NYC and Frankfurt crawls differ slightly in the number of ads: we scraped 
10,278 ads in the NYC crawl and only 9,135 in the Frankfurt crawl---the latter is the crawl with the least amount of ads. 
Slightly more (29 vs 26\%) websites in the NYC mobile crawl have targeted ads;
and NYC mobile crawl has the highest proportion of targeted ads (82\%) across all crawls. We also discovered that improper ads, particularly racy and weight loss ads, were more prevalent on mobile devices compared to desktops.

\vspace{-0.15cm} 
\section{Discussion}
Our research paints a troubling picture of tracking and inappropriate advertising practices on child-directed websites. Advertisements featuring sexually suggestive imagery and ads about weight loss, dating, and mental health may pose potential risks to children's emotional and psychological welfare. We discuss the legal implications, ethical considerations and limitations of our study below.

\subsection{Legal implications}
\label{subsec:legal}
In this section, we discuss what the law says about tracking and advertising practices uncovered in our research.
We focus on the EU General Data Protection Regulation (GDPR) and the US Children's Online Privacy Protection Act (COPPA)\footnote{We do not analyze whether specific companies breach the law. For such an analysis, each case would have to be examined separately, considering all the circumstances of that specific case. Rather, we discuss legal requirements in general terms.}.

\textbf{The GDPR and the ePrivacy Directive.}
Under the GDPR, companies are only allowed to process personal data if they have a `legal basis' for such processing. The GDPR provides six possible legal bases (article 6 GDPR). However, generally, the data subject's consent is the only possible legal basis for online tracking and behavioral (targeted) advertising~\cite{european2022guidelines}. Moreover, the ePrivacy Directive~\cite{ePrivacyDirective2009} requires, in short, companies to ask the internet user for consent before they use tracking cookies or similar tracking technologies (Article 5(3)). 

The GDPR's requirements for valid consent are strict. Consent is only valid if it is really voluntary (`freely given'), and `specific' and `informed'.
The data controllers (the website owner and the companies involved in tracking and targeted advertising) must `be able to demonstrate that the data subject has consented to process of his or her personal data' (Article 7(1) GDPR). The GDPR's requirements for valid consent also apply to consent (for cookies etc.) as prescribed by the ePrivacy Directive. 
The GDPR has specific rules for consent by children. Roughly summarized, children cannot give valid consent; the parent should give consent instead (Article 8 GDPR). EU member states have set different minimum consent ages, ranging from 13 to 16 years~\cite{milkaite2019status}. Hence, only parental consent can legitimize tracking on a children's website.
Observe that a parent clicking a consent dialog (as done by our crawler) does not constitute parental consent under GDPR. Even in low-risk cases, verification of parental responsibility via email may be necessary~\cite{european2020guidelines}.

\textbf{The EU Digital Services Act.}
The rules for tracking and targeting children will become stricter in the EU. From 17 February 2024 on, the EU Digital Services Act~\cite{DSA2022} applies. Article 28 says, roughly summarized, that online platforms must not use behavioral advertising `when they are aware with reasonable certainty that the recipient of the service is a minor'~\cite{DSA2022}. This prohibition cannot be overridden with the consent of the child or the parent. 
The DSA also requires ``very large online platforms''~\cite{EC_VLOP_List} (with more than 45 million users in the EU) to publish the advertisements that it presented to users in an online repository, together with information about, for instance, the targeting criteria (Article 33, 39 DSA). The methods that we used in this paper could be used to check the completeness and accuracy of data published in those repositories.      

\textbf{COPPA.}
COPPA regulates companies offering a website or online service directed to children under the age of 13. Specifically, COPPA applies to companies using children's `personal information,' which includes `persistent identifiers such as cookies and device fingerprints' (COPPA \S312.2)~\cite{FTC2020}. The website owner is responsible for data collection by third parties through its site. Such third parties must also comply with COPPA. Companies based outside the US must also comply with COPPA if their services are directed to children in the US~\cite{FTC2020}.

Our results showed that 27\% of the child-directed websites use targeted advertising.
Under COPPA, data collection for targeted advertising on these websites is only allowed after getting parents’ Verifiable Parental Consent (VPC).
VPC entails utilizing stringent verification methods, such as credit card verification, face recognition, or government ID checks~\cite{FTC2022}.
This makes VPC much more complex than simply clicking an accept button on a dialog. We note that our crawler cannot simply give VPC. 

\subsection{Research Ethics}\label{research-ethics}
Our crawler visited over 166K pages and it triggered many ad impressions that could be viewed by a real visitor (likely a child). 
Given the huge scale of the digital ad market (projected to reach US\$700bn in 2023~\cite{statista-global-admarket}) 
we believe these ad impressions are a negligible cost for raising the transparency around tracking and ads targeted to children. 
Furthermore, we took several measures to limit our footprint on the crawled websites. For instance, we only crawled five inner pages from each site in a crawl, and we randomly shuffled the target URLs to avoid concurrently visiting the inner pages of a website. 
We also took appropriate measures to ensure that no harm was done to collaborators involved in the project, especially when dealing with explicitly graphic images. 

\textbf{Disclosures and outreach}
In July 2023, we reached out to five companies that we found to serve racy ads. 
One company invited us to a call and explained the likely reason for showing inappropriate ads (websites failing to label themselves as child-directed and a false negative in the ad company's automated child-directed site detection tool). They added the sites in question to the child-directed list and pledged further investigation. Another thanked us and began an internal review. The third redirected our query to the relevant department.
Moreover, we disclosed 34 racy ads to Google by manually visiting the ad disclosure URLs of each racy (Google) ad; and using the \textit{Report this ad} button.
Note that, to identify the ad vendors involved in serving the ad, we used a combination of ad images, and src/href attributes of the ads' descendant iframe, image and link elements (\S\ref{ad_detection}).

In addition, we shared our preliminary results with a European data protection agency (DPA), and a consumer protection agency.
Both showed interest; the DPA asked if there were any websites from their country containing improper ads.
The consumer protection agency invited us to present our work.
We also shared our findings with civil society and industry organizations including the 5Rights Foundation~\cite{5Rights}.
We plan to further share our study with regulators and other relevant stakeholders.

While using the VirusTotal API, we found and reported three porn websites miscategorized as kids-related to the respective third-party categorization service. Although no response was received, the categories were later rectified.

\subsection{Limitations}\label{limitations-future-work}
Our study predominantly covers websites targeting younger children, as we define `children' as under 13, aligning with both US and EU regulations. While our classifier detected child-directed sites in 48 languages, it may have a potential bias towards English content, which is overrepresented in the training data. The classifier may prefer sites with descriptive titles and content or may carry biases related to website age, design, or accessibility.
Moreover, the classifier favors precision over recall to reduce the manual labeling workload.

Our research is the first attempt to build a large list of child-directed websites.
Classifying websites as child-directed or not is challenging, primarily due to the existence of gray areas that complicate the labeling process.
For instance, many websites offer content that appeals to both children and adults.
While we tried to exclude websites that can only be used by teachers or parents, certain websites included pages such as online exercises, videos or games that can be consumed by children.
To validate our list, we performed further analysis on two subsets of websites: two senior researchers relabeled a random 100-website sample; one of the senior researchers relabeled a random subset of 100 websites with targeted (50) and inappropriate ads (50). In both cases, we identified less than 9\% of websites that are related to children, but mainly catered to teachers or parents (8.6\%, 8.3\%, respectively). 
While this impurity could be avoided with more conservative labeling, our analysis of 200 websites strongly suggests these cases do not skew our primary findings.

While we found fewer targeted ads in the EU than in the US,
we cannot directly attribute this to differences in privacy regulation or another specific factor.
Failure to detect and interact with consent dialogs may be a confounding factor, among others.

When detecting targeted ads, we only used ad disclosure pages from two providers (Google and Criteo) due to the unavailability of useful ad disclosures from other vendors. Thus, our targeted ad detection method depends on the accuracy, completeness and precision of Google and Criteo's ad disclosures.

Websites may treat cloud-based IP addresses or automated browsers differently~\cite{zeber2020representativeness,jueckstock2021towards,cassel2022omnicrawl}. To curb such effects, we used the anti-bot detection features of TRC~\cite{duckduckgo-trc}.
Reviewing the screenshots captured during the visits, we observed very few blocked visits.

We conducted four sets of inner link collection crawls: two from NYC and two from Frankfurt, encompassing both desktop and mobile crawls. 
This constraint does not appear to impact the success rate of visits across these vantage points; nonetheless, future research could explore the possibility of identifying inner pages during the crawling process.

Since we use a fresh profile for each page visit, we may not capture re-targeted or other personalized ads that are only shown to users with a behavioral profile. Future work could extend our method to incorporate personas and warm-up crawls to study such ads.
Overall we do not claim that our findings are representative of tracking and advertising practices on child-directed websites.
Our focus in this study is not on how ads are targeted, but simply on \textit{whether the targeting is enabled or not}.

\vspace{-0.2cm}

\section{Conclusion}
We presented an empirical study of online tracking and advertisements on over 2,000 child-directed websites.
Building a lightweight and versatile ML pipeline to analyze ad content, we identify hundreds of cases of improper ads, including weight loss and mental health ads, and ads featuring dating services, racy and sexually suggestive imagery.
Our study reveals several notable trends: websites featuring advertisements tend to contain two to four times more trackers, mobile websites exhibit a greater prevalence of inappropriate ads, and popular websites are less likely to deploy targeted advertisements.
Our findings provide concrete evidence of troublesome practices that are likely illegal, unethical, or simply careless. 
We call for more research, regulation and enforcement to limit the ongoing violation of children's privacy and well-being. 

\section{Acknowledgments}
Asuman Senol was funded by the Cyber-Defence (CYD) Campus of armasuisse Science and Technology.
Veelasha Moonsamy was funded by the Deutsche Forschungsgemeinschaft (DFG, German Research Foundation) under Germany’s Excellence Strategy - EXC 2092 CASA - 390781972.

\bibliographystyle{IEEEtran}
\bibliography{ref}

\appendices
\section{Criteria for labeling child-directed websites}\label{criterion}

To identify a child-directed website, we manually visit and review its design, content, and policies, including necessary translations to English. A site is labeled as child-directed if any of the following conditions are met:
\begin{itemize}
    \item Does the website include content, activities, or games that can be used by children?
    \item Does the website promote products (e.g., apps, sites, books, videos, workshops, animations, etc.) designed for and usable by children online?
    \item Does the website include content or promote products whose end users are children, but children's parents must first subscribe or register?
\end{itemize}
A site is not child-directed if one of the following is true:
\begin{itemize}
    \item The website redirects to another page that is not child-directed.
    \item The website features children-related products intended for adult use, such as parents or teachers.
    \item The website is generally appealing to adults (e.g., news or academic websites).
\end{itemize}

\subsection{Manual Verification of the Classifier Output}\label{manual-verif}

Two researchers manually labeled a total of 2,500 websites detected as child-directed by the classifier. Initially, both researchers jointly labeled a 50-website sample, reaching agreement on 45 of these decisions (Cohen’s Kappa=0.79~\cite{cohen-kappa}). The remaining websites were divided into two equal batches and labeled individually by the two researchers. This process took approximately one person-week to complete.
We followed the criteria for identifying child-directed websites (Appendix~\ref{criterion}) and considered four potential labels for each website: \textit{child-directed}, \textit{child-related}, \textit{non-children} and \textit{unknown}.
The majority of the websites (64\%) were labeled as child-directed, 
, while 23\% were identified as child-related, indicating they are relevant to children but primarily intended for parents or teachers. About 4.5\% of the websites were inaccurately classified as non-children's sites, such as entertainment bands, academic forums, and movie streaming services, mostly due to vague or brief titles and descriptions.
In some cases, discerning whether websites were targeted at children, parents, or teachers was challenging, leading to 5.5\% being labeled as `unknown' due to uncertainty about their target audience. Of the misclassified websites, we found that four were adult entertainment websites (0.16\%) that had very short metadata fields mentioning words such as ``teens'', ``cartoon'', ``animations'' which likely caused the misclassification. 

\subsection{Ad Transparency Statements}\label{ad_labeling}

The following ad transparency statements are used to classify advertisements as targeted or non-targeted. Note that targeted categories also include retargeting and behavioral ads. The statements are compiled from Google's and Criteo's ad disclosure interfaces, reached via the AdChoices icon. 

When searching for the statements, we use exact, case-insensitive search.

\textbf{Targeted:}
\begin{itemize}
    \item Google's estimation of your interests
    \item Websites you've visited
    \item Your similarity to groups of people the advertiser is trying to reach
    \item Your activity on Google on this device
    \item According to your activity on this device
    \item You have enabled ad personalization
    \item Information collected by the publisher. The publisher partners with Google to show ads
    \item Google's estimation of the languages you know, based on your activity on this device
    \item Your visit to the advertiser's website or app
    \item The advertiser's interest in reaching new customers who haven't bought something from them before
\end{itemize}

\textbf{Non-Targeted:}
\begin{itemize}
    \item Ad personalization is turned off
    \item You have turned off ad personalization
    \item Ad personalization is off
    \item The time of day or your general location
    \item Google shows ads based on general factors like the time of day and the info on a page, our policies, and your ad personalization settings
    \item The information on the website you were viewing
    \item General factors about the placement of the ad
\end{itemize}

\subsection{Detecting failed or errored visits}\label{failed-visits}
We classified a visit as failed under the following conditions: if the first request elicited a 4XX or 5XX error; if the size of the first non-3XX response (root document) was less than 512 bytes, following Le Pochat et al.~\cite{tranco_list}; or if there was no successful (200 OK) response.

\end{document}